\documentclass{article}
\usepackage[T1]{fontenc}
\usepackage{wrapfig}
\usepackage[portuges,english]{babel}
\usepackage{amssymb,latexsym,amsmath,color,mathrsfs,ifsym,graphics,stmaryrd} 
\usepackage[colorlinks,linkcolor=blue,urlcolor=blue,citecolor=black,
plainpages=false,pdfpagelabels,breaklinks]{hyperref}
\usepackage{dialogue}
\usepackage{graphicx}
\title{{\sc Bohr's Anti-Realist Realism in Contemporary\\(Quantum) Physics and Philosophy.}}

\author{{\sc Christian de Ronde}\thanks{Fellow Researcher of the Consejo
Nacional de Investigaciones Cient\'{\i}ficas y T\'ecnicas. E-mail: cderonde@gmail.com}}
\date{}

\usepackage[margin=2.15cm]{geometry}

\begin{document}
\maketitle
\begin{center}
\begin{small}
Philosophy Institute Dr. A. Korn, Buenos Aires University - CONICET\\ 
Engineering Institute - National University Arturo Jauretche, Argentina.\\
Federal University of Santa Catarina, Brazil.\\ 
Center Leo Apostel for Interdisciplinary Studies, Brussels Free University, Belgium.\\
\end{small}
\end{center}

\bigskip 

\begin{abstract}
\noindent We discuss the influential role of Niels Bohr's work in the anti-realist realist re-foundation of physics that took place during the 20th century. We will focus in how, developing the modern co-relational matrix of scientific understanding, his essentially anti-realist scheme was able to capture, subvert and defeat the realist program of science through the establishment of a weakened impotent form of ``religious realism'' grounded on faith instead of scientific conditions. Finally, we will focus in how, still today,  anti-realist realism continues to rule the contemporary post-modern research in both (quantum) physics and philosophy. 

\medskip
\noindent \textbf{Key-words}: Realism, anti-realism, representation, quantum theory.
\end{abstract}

\renewenvironment{enumerate}{\begin{list}{}{\rm \labelwidth 0mm
\leftmargin 0mm}} {\end{list}}

\newcommand{\ita}{\textit}
\newcommand{\mcal}{\mathcal}
\newcommand{\mfrak}{\mathfrak}
\newcommand{\mbb}{\mathbb}
\newcommand{\mrm}{\mathrm}
\newcommand{\msf}{\mathsf}
\newcommand{\mscr}{\mathscr}
\newcommand{\lra}{\leftrightarrow}
\renewenvironment{enumerate}{\begin{list}{}{\rm \labelwidth 0mm
\leftmargin 5mm}} {\end{list}}

\newtheorem{dfn}{\sc{Definition}}[section]
\newtheorem{thm}{\sc{Theorem}}[section]
\newtheorem{lem}{\sc{Lemma}}[section]
\newtheorem{cor}[thm]{\sc{Corollary}}
\newcommand{\Proof}{\textit{Proof:} \,}
\newcommand{\cqd}{{\rule{.70ex}{2ex}} \medskip}

\bigskip

\bigskip

\bigskip

\begin{flushright}
{\small {\it The metaphysicians of Tl\"on do not seek for the truth\\
or even for verisimilitude, but rather for the astounding.\\ 
They judge that metaphysics is a branch of fantastic literature. }\\
\smallskip
Jorge Luis Borges} 
\end{flushright}


\section*{Introduction}

Since its Greek origin, physics has been related to {\it physis}, namely, the totality of what {\it is}. The realist presupposition that gave birth to physics was the idea that {\it theories} provide knowledge about the {\it logos} (i.e., order) of reality through the creation of systematic, unified schemes capable to account for the multiplicity immanently found within experience (see for discussion \cite{Cordero05, Cordero14}). This was the case for more than two millennia of successful developments from Protagoras and Heraclitus to Plato and Aristotle, and then, up to modern times to Galileo and Newton. But even though modernity ---with the creation of classical mechanics--- could be regarded as the peak of the Greek theoretical realist program, this period can be also seen as the starting point of the anti-realist re-foundation of science. A process that would culminate in post-modern times, during the 20th century. As Karl Popper would famously describe the situation during the late 1950s:  
\begin{quotation}
\noindent {\small ``Today the view of physical science founded by Osiander, Cardinal Bellarmino, and Bishop Berkeley, has won the battle without another shot being fired. Without any further debate over the philosophical issue, without producing any new argument, the {\it instrumentalist} view (as I shall call it) has become an accepted dogma. It may well now be called the `official view' of physical theory since it is accepted by most of our leading theorists of physics (although neither by Einstein nor by Schr\"odinger). And it has become part of the current teaching of physics.''  \cite[pp. 99-100]{Popper63}}
\end{quotation}
Physical theories would then become to be regarded as an economy of `clicks' in detectors not necessarily linked to the description of reality. Essential in order to understand this period of radical transformation is to pay special attention to what happened, during the 20th century, in a very specific field of scientific research that would become the central arena of battle between realists and anti-realists, namely, the theory of quanta. Popper \cite[p. 100]{Popper63} himself would link the reasons behind the triumph of anti-realism to Quantum Mechanics (QM): ``How then did [instrumentalism] come about? As far as I can see, through the coincidence of two factors, (a) difficulties in the interpretation of the formalism of the Quantum Theory, and (b) the spectacular practical success of its applications.'' According to Popper while (a) was intrinsically related to Bohr's complementarity interpretation of QM, (b) was justified by the {\it Manhattan Project} which culminated with a ``big bang''. Although we agree with these points, we believe there is a kernel element ---not mentioned by Popper--- which is essential in order to understand the effectiveness behind the anti-realist victory. As we will argue in this work, the success of anti-realism during the 20th century was not grounded on the eradication of realists but on their control and effective conversion to the dogmas imposed by their colonizers. As a consequence, the new generations of brainwashed realists would learn to accept not only their own impotency to solve anything but also their pathetic role as ``fanatic believers'' in the capacity of mythical narratives to unveil reality-in-itself ---regardless of any scientific experimental or theoretical support.  

Today, after  Bohr's quantum revolution, physics has become flexible enough to contain not only a purely pragmatic (anti-realist) account of the discipline in terms of models designed to solve technical problems but also, a purely abstract (supposedly realist) mathematical account of physics which searches a formalistic unification that will, in turn, lead to the final discovery of the ``Theory Of Everything'' (T.O.E.) \cite{Weinberg93}. Of course, in the meantime, questions made by students about the reality of quanta continue to be standardly responded by Professors in classrooms all around the world exactly with the same mantra: ``Shut up and calculate!'' Quite recently, Sean Carroll \cite{Carroll20} has revealed what happens to the few students which refuse to follow such un-scientific orders: ``Many people are bothered when they are students and they first hear [about Standard Quantum Mechanics]. And when they ask questions they are told to shut up. And if they keep asking they are asked to leave the field of physics.'' Fortunately enough, since the 1980s some of these outcasts with obvious realist inclinations have been embraced by a new field outside the gates of instrumentalism, namely, philosophy of quantum mechanics. Unfortunately, when analyzed in detail one might reach the conclusion that this new field of research ---created by anti-realists themselves--- has not encouraged the critical consideration of the empiricist re-foundation of science but, on the very contrary has helped to shield the anti-realist understanding of empirical science itself. In fact, this concentration camp has become essential for anti-realists to control the activities, problems and debates discussed by realists rebels who, step by step, have been forced to convert themselves to the new creed. It is in this context that, closer to fanatic believers of a religious sect than to rational scientists, contemporary realists have come to consent their sad role as buffoons of the anti-realist Court. 

According to the anti-realist characterization of realism ---which realists themselves have learned to accept---, realists are naive subjects who believe that their made up narratives ---called by philosophers ``interpretations''--- have the force to provide a true correspondence description of reality-in-itself. Quite regardless of the complete lack of any (objective) link to the theories beyond aesthetic or voluntarist choices \cite{Chakravartty17, French11, Raoni20}, such fictional interpretations have played an essential role helping many troubled researchers to replace their metaphysical anxieties with the faith that their interpretations do provide a true access to the world ---going in this way beyond the pragmatic models developed by contemporary empirical science. Sadly enough, physicists who seldom pay any attention to the efforts of realist philosophers regard most of these stories as lying beyond the scope of science.\footnote{As Maximilian Schlosshauer \cite[p. 59]{Schlosshauer11} has recently described: ``It is no secret that a shut-up-and-calculate mentality pervades classrooms everywhere. How many physics students will ever hear their professor mention that there's such a queer thing as different interpretations of the very theory they're learning about? I have no representative data to answer this question, but I suspect the percentage of such students would hardly exceed the single-digit range.''} And of course, since realists have no rational justification for their professed faith in the interpretation of their choice, they are essentially correct. This has become explicitly clear in the context of QM where the number of interpretations has created a ridiculous situation accurately described by David Mermin \cite[p. 8]{Mermin12}: ``[Q]uantum theory is the most useful and powerful theory physicists have ever devised. Yet today, nearly 90 years after its formulation, disagreement about the meaning of the theory is stronger than ever. New interpretations appear every day. None ever disappear.'' Indeed, the ``realist creation of fictional narratives'' seems to have gone completely out control in philosophical journals, generating what \'Adan Cabello has termed ``a map of [interpretational] madness'' \cite{Cabello17}. It is in this context that it becomes clear that realist believers have ended up only making fun of themselves helping in this way ---willingly or not--- to actually reinforce the anti-realist {\it status quo} and the claim that interpretations are just human hallucinations which unfruitfully attempt to reach the unreachable. 

To conclude, anti-realists have applied in a very efficient way the strategy proposed by Michael Corleone in {\it The Godfather}: ``Keep your friends close, but your enemies closer.'' Confined to a Promethean task realists have been forced to endlessly create un-useful fictions which have no direct link to the formalism they attempt to represent nor to the experience they are supposed to explain. Imprisoned in this labyrinth, they wonder as fanatic preachers trying to sell their made up stories in philosophical journals. For the first time in the history of a war which begun more than two and a half millennia ago, anti-realists have been able to capture realists in their rhetorical net and conquer the totality of contemporary knowledge. However, even though the anti-realist illusion has remained extremely effective, like in every majic trick the deception is accomplished only if the spectator uncritically believes everything she has been told by the conjurer is actually true.


\section{What Was the (Pre-Bohrian) Role of Realism Within Physics?}

Realism is certainly not ---as it has been cartoonized by anti-realsits--- the activity of ``interpreting'' mathematical algorithms through the invention of narratives in order to ---later on--- ``fanatically believe'' they are true descriptions of ---what Kant called--- the {\it Thing-in-Itself}. It is neither the attempt to describe the presupposed entities observable or not which constitute the ontological ``furniture of the world'' or ``the stuff the world is made off''. The true realist {\it praxis} has been always related to the theoretical creation of formal-conceptual {\it moments of unity} capable to account in a coherent and consistent manner for the multiplicity found in different fields of experience ---independently, of course, of subjective viewpoints. As Wolfgang Pauli would explain to a young Werner Heisenberg:   
\begin{quotation}
\noindent {\small  ``[...] knowledge cannot be gained by understanding an isolated phenomenon or a single group of phenomena, even if one discovers some order in them. It comes from the recognition that a wealth of experiential facts are interconnected and can therefore be reduced to a common principle. [...] `Understanding' probably means nothing more than having whatever ideas and concepts are needed to recognize that a great many different phenomena are part of coherent whole. Our mind becomes less puzzled once we have recognized that a special, apparently confused situation is merely a special case of something wider, that as a result it can be formulated much more simply. The reduction of a colorful variety of phenomena to a general and simple principle, or, as the Greeks would have put it, the reduction of the many to the one, is precisely what we mean by `understanding'. The ability to predict is often the consequence of understanding, of having the right concepts, but is not identical with `understanding'.'' \cite[p. 63]{Heis71}}
\end{quotation}
For this reason, the only foundation that can be accepted by realists is {\it physis} itself; in Parmenidean terms, the one, everything which {\it is} \cite{Cordero05}. As Heraclitus would declare: ``Listening not to me but to the {\it logos} it is wise to agree that all things are one'' [f. 50 DK]. From a realist perspective, in the case of physics, and apart from physis, the {\it moments of unity} considered {\it within} reality are theoretical creations, formal-conceptual nets weaved immanently in order to express the multiple relations found within a field of experience. Instead of a transcendent description we are talking here of an immanent expressivity. In this way, theories are capable of expressing an immanent order (or {\it logos}), capturing singular truths within the infinite set of relations that constitute the totality of being. As Jorge Luis Borges exposed in his famous story, {\it Funes the Memorious}, moments of unity do not appear when we open our eyes, for the unity of experience is not something {\it given} to our senses. Following Immanuel Kant the categorical conditions of experience (e.g., existence, no-contradiction, identity, causality, etc) are not principles that can be ``found'' through observation but metaphysical presuppositions which allows us to make sense of what is being observed in terms of a unity. Consequently, reality ---at least for the realist--- can never be understood as something composed of ``individual physical entities'' which are waiting to be unveiled through theories, models and interpretations. Theories ---in the realist sense--- have never attempted to mirror or bring into presence ``the stuff the world is made off'' or ``the furniture of the world''. This is, in fact, a ``common sense'' ontological presupposition that comes from anti-realism. On the very contrary, the Greek program of science can be regarded as a continuous attempt to escape ``common sense'' and create always new theoretical {\it moments of unity} which allow us to think in coherent and consistent terms (i.e., rationally) about experience independently of subjective viewpoints; or, to say it in modern physical terms, independently of reference frames. Only when discussing about something explicitly defined and common to all speakers rational scientific discourse and debate become possible. In search for the relation between the one and the many, it was the Ancient Greek problem of movement which guided physics and philosophy for more than two millennia: How can {\it the same} be considered as {\it changing}? What is {\it identity} within {\it difference}? After the deep criticism of sophistry during the Vth and IVth centuries B.C. the realist program of science undertook with Plato's {\it Sophist} an essential metaphysical turn, creating the first categorical systematization of being in terms of {\it dynamis}, i.e., relations. Some decades later, it would be one of his students, Aristotle, who would generate another powerful metaphysical system that would consider the movement of {\it entities} as the path from a potential mode of existence to an actual one categorically defined in terms of the principles of existence, non-contradiction and identity ---giving in this way also rise to classical logic itself. However, it would be only after two millennia, in modern times, that the Greek program of science would reach its peak with the development of the first theory capable to unite terrestrial and celestial phenomena. Essential to this development are two notions that expose the entangled relation between mathematical formalisms and conceptual (or metaphysical) systems, namely, mathematical invariance and conceptual objectivity. Indeed, during modernity, the Greek quest was specified in the following terms: What can be regarded to be {\it the same} through {\it change} in objective and invariant terms? Let us explain these kernel notions of modern physics in some detail.



As it is well known, physical representation presupposes the choice of what is called a `reference frame', a `mathematical perspective' which is considered to be a pre-requisite for any meaningful description of a state of affairs. A system can be defined in abstract terms through its properties and their values.\footnote{Classical mechanics and later on the theory of electromagnetism constructed their own specific understanding of what a system {\it is}. While in classical mechanics the system would be described in terms of the notion of `particle' or `rigid body' (i.e., a compound of particles), in electromagnetism the main notion would become that of `wave'.} But even though there exist some properties which will not vary when considered from different reference frames (i.e., invariant properties such as the mass or the charge of a particle) there are others which will possess values only {\it relative} to a frame of reference (e.g., position and velocity). Thus, the representation provided by each distinct reference frame will in general differ. The particle will have position $x_1$ and velocity  $v_1$ {\it relative}  to the frame $R_1$ but $x_2$ and  $v_2$  {\it relative}  to $R_2$. It is the physical notion of {\it state} which has the main purpose not only to provide the specific values of the properties of the system within a specific situation but also to consistently unite all different viewpoints in terms of {\it sameness}. It is the consistent translation between different reference frames given through invariant transformations\footnote{Mathematical invariance is a very general abstract concept which can be defined in the following manner. Let $G$ be a group acting on spaces $V$ and $W$. A function $f:V\to W$ is  called \emph{invariant} if $f$ maps orbits to orbits, that is,  if $f(g\cdot v)=g\cdot f(v)$ for all $g\in G$ and $v\in V$.} which allows us to provide a {\it global} consistent account of {\it the same} state of affairs as completely detached from reference frames themselves. This is the key to make sense of the idea that different empirical observers can have different descriptions of a situation and yet be talking about the same state of affairs. It is this kernel consistency which, in fact, allows us to talk about `something' independent of empirical perspectives. Obviously, if different observers would disagree about a {\it common} account of the same state of affairs, then the realist program of science would become simply impracticable ---as well as rational discourse. 

The idea that different reference-frame dependent representations should describe {\it the same} was ---in fact--- the thread of Ariadna which guided Einstein in the development of his special theory of relativity. In this case, it was not the ``common sense'' understanding of space and time phenomena inherited from Newtonian mechanics but the {\it principle of relativity} ---which pointed explicitly to the operational invariance requested to consider experiments in different reference frames--- which ``had to be saved''. According to it, there must exist an equivalence relation between the descriptions provided in different inertial reference frames, even if that means to leave behind the modern representation of space and time as being the absolute containers of the physical world. In this respect, the main difference between classical mechanics and Einstein's relativity is that while in the first case spatial and temporal intervals are considered as invariant with respect to different reference frames and the velocity of light changed according to the Galilean transformations, in the latter case exactly the opposite becomes true: while temporal and spatial intervals are dependent on reference frames the speed of light becomes an invariant of the theory. While in the first case it is the Galilean transformations which provide global consistency to positions and velocities of particles, in the latter case it is the Lorentz transformations that secure the global consistency of spatial and temporal intervals as invariant variations of the theory. 

In short, the realist quest attempts to create in a systematic manner (meta-)physical concepts which generate {\it moments of unity} allowing us thus to conceive a specific field of experience described from {\it different} (formal-conceptual) perspectives in terms of {\it the same} state of affairs. Einstein would stress this point repeatedly:  
\begin{quotation}
\noindent {\small``From Hume Kant had learned that there are concepts (as, for example, that of causal connection), which play a dominating role in our thinking, and which, nevertheless, can not be deduced by means of a logical process from the empirically given (a fact which several empiricists recognize, it is true, but seem always again to forget). What justifies the use of such concepts? Suppose he had replied in this sense: Thinking is necessary in order to understand the empirically given, {\it and concepts and `categories' are necessary as indispensable elements of thinking.}'' \cite[p. 678]{Einstein65} (emphasis in the original)}
\end{quotation} 
He would also warn physicists of the dangers of taking concepts as natural or ``common sense'' {\it givens}. 
\begin{quote}
``Concepts that have proven useful in ordering things easily achieve such an authority over us that we forget their earthly origins and accept them as unalterable givens. Thus they come to be stamped as `necessities of thought,' `a priori givens,' etc. The path of scientific advance is often made impossible for a long time through such errors.'' \cite[p. 102]{Howard10}
\end{quote}
For the realists, as Einstein would tell Heisenberg \cite{Heis71}: ``it is only the theory which can tell you what can be observed.'' An essential point also stressed by Heisenberg in his elder works: 
\begin{quote}
``The history of physics is not only a sequence of experimental discoveries and observations, followed by their mathematical description; it is also a history of concepts. For an understanding of the phenomena the first condition is the introduction of adequate concepts. Only with the help of correct concepts can we really know what has been observed.''   \cite[p. 264]{Heis73}
\end{quote}
Consequently, observation ---unlike for the anti-realist--- does not provide a direct access to ``external reality'', but ---on the very contrary--- is a procedure {\it derived} from theoretical considerations. This metaphysical standpoint marks the crossroad between realism and anti-realism. While for anti-realist the existence of individual subjects and objects appears as a ``common sense'' standpoint of analysis, for realists {\it moments of unity} are always created theoretically, i.e., in a formal-conceptual systematic manner. Theories do not describe or unveil ``the furniture of the world'' but immanently express the order that can be found and constructed between relations. The essential {\it praxis} of realists involves the attempt to build unity and sameness in order to organize the multiplicity found within experience. Realists are essentially theory constructors, creators of formal-conceptual moments of unity (e.g., physical systems) which must always remain independent of reference frames (i.e., states) in order for us to conceive reality (immanently) in terms of an evolving objective-invariant state of affairs. As we will see in the following section, it is this very general realist scheme which was effectively subverted by the most influential physicist of the 20th century, the Danish Niels Bohr.

\section{Bohr's Anti-Realist Realism}

What is real? This question has a simple answer for the realist: it is only reality which can be regarded as real. So what is realty? According to realism it is everything which {\it is}, the totality of what can {\it be}, what the Greeks called {\it physis}. This is, in fact, the only direct ---obviously presupposed--- reference of realism which blocks ---as Parmenides argued--- the possibility of {\it non-being}, of nothingness entering or cutting the realm of existence. Reality ---contrary to anti-realists--- cannot be separated in realms, it is an encompassing whole, it is one. As a second element of the realist program we find the idea that {\it theories} ---which are of course part of reality, as everything else--- are capable of relating to {\it physis} in scientific terms providing an immanent form of knowledge.\footnote{How this is actually done is of course a problematic subject for the realist, specially after Kant's correlationalism.} However, theories cannot be regarded as real in the same sense as {\it physis} is real, they do not mirror or unveil a reality composed of `things' or `stuff' ---as the ontological picture of anti-realism presupposes. Theories are immanent nets designed to capture a specific field of phenomena, through the weaving of formal and conceptual relations, true expressions of reality. In contraposition to this line of reasoning, today, anti-realist have encouraged a {\it praxis} according to which the term `real' can be stamped to anything: observations, objects, events, mathematical entities, equations, `clicks' in detectors, minds, fields, thoughts, particles, waves, consciousness, interpretations, narratives, models or even theories themselves. This might lead us to the wrong conclusion that we are witnessing an historical period where the ruling of realism continues to guide the fate of Western thought as it did, since the Ancient Greeks, for more than two millennia. On the very contrary, this situation is the result of the dissection of reality, the subversion of its meaning followed by its complete fragmentation into nothingness. The destruction of reality is a direct consequence of the undisputed triumph of anti-realism over realism. A victory that was constructed since modernity and Enlightenment through two essential displacements of the basic realist cornerstones. First, the change of perspective from {\it physis} to the subject, and second, the carefully separation and division of reality ---and knowledge--- into smaller and smaller binary compartments. {\it Divide et Impera}. This is the motto which guided anti-realism successfully in its quest to destroy the essential reference of her enemy. First Ren\'e Descartes and then Immanuel Kant would effectively cut into pieces the Greek concept of reality. During the 16th. century {\it physis} would be first separated by the French philosopher in three distinct realms: {\it res cogitans}, {\it res extensa} and God. Something precisely explained by Werner Heisenberg: 
\begin{quotation}
\noindent {\small ``The great development of natural science since the sixteenth and seventeenth centuries was preceded and accompanied by a development of philosophical ideas which were closely connected with the fundamental concepts of science. It may therefore be instructive to comment on these ideas from the position that has finally been reached by modern science in our time.

The first great philosopher of this new period of science was Ren\'e Descartes who lived in the first half of the seventeenth century. Those of his ideas that are most important for the development of scientific thinking are contained in his {\it Discourse on Method}. On the basis of doubt and logical reasoning he tries to find a completely new and as he thinks solid ground for a philosophical system. He does not accept revelation as such a basis nor does he want to accept uncritically what is perceived by the senses. So he starts with his method of doubt. He casts his doubt upon that which our senses tell us about the results of our reasoning and finally he arrives at his famous sentence: {\it `cogito ergo sum.'} I cannot doubt my existence since it follows from the fact that I am thinking. After establishing the existence of the I in this way he proceeds to prove the existence of God essentially on the lines of scholastic philosophy. Finally the existence of the world follows from the fact that God had given me a strong inclination to believe in the existence of the world, and it is simply impossible that God should have deceived me.

This basis of the philosophy of Descartes is radically different from that of the ancient Greek philosophers. Here the starting point is not a fundamental principle or substance, but the attempt of a fundamental knowledge. And Descartes realizes that what we know about our mind is more certain than what we know about the outer world. But already his starting point with the `triangle' God-World-I simplifies in a dangerous way the basis for further reasoning. The division between matter and mind or between soul and body, which had started in Plato's philosophy, is now complete. God is separated both from the I and from the world. God in fact is raised so high above the world and men that He finally appears in the philosophy of Descartes only as a common point of reference that establishes the relation between the I and the world.

While ancient Greek philosophy had tried to find order in the infinite variety of things and events by looking for some fundamental unifying principle, Descartes tries to establish the order through some fundamental division. But the three parts which result from the division lose some of their essence when any one part is considered as separated from the other two parts. If one uses the fundamental concepts of Descartes at all, it is essential that God is in the world and in the I and it is also essential that the I cannot be really separated from the world. Of course Descartes knew the undisputable necessity of the connection, but philosophy and natural science in the following period developed on the basis of the polarity between the {\it `res cogitans'} and the {\it `res extensa,'} and natural science concentrated its interest on the {\it `res extensa.'} The influence of the Cartesian division on human thought in the following centuries can hardly be overestimated, but it is just this division which we have to criticize later from the development of physics in our time.'' \cite[pp. 41-42]{Heis58}} 
\end{quotation}  
Following Descartes, Kant's co-relational scheme would replace {\it res cogitans} with the {\it transcendental subject}, {\it res extensa} with {\it objects of experience} and God with a newly designed concept: {\it the-thing-in-itself}.\footnote{A notion which marks the essential (anti-realist) re-direction of science from the notion of {\it physis} to that of `things'. Something that would be developed, during that same period, in terms of the notion of ontology. According to the Oxford Dictionary \cite{OxfordDictionary} in Philosophy: ``Ontology is the study or concern about what kinds of things exist ---what entities or `things' there are in the universe.''} But while in the case of Descartes it would be God himself who would serve as a warrant for securing the truthful relation between the ``internal'' cognitive reality of the subject and the ``external'' material reality of objects, in the case of Kant this bridge would be built in purely {\it a priori} terms, leaving reality-in-itself as a mere regulatory ideal, a ghostly shadow which, as remarked by Friedrich Jacobi [1787: 223] stood in a paradoxical situation, inside and outside the whole system: ``Without the presupposition [of `the-thing-in-itself,'] I was unable to enter into [Kant's] system, but with it I was unable to stay within it.'' Kant, a physicist himself, had finally detached science from {\it physis} and redirected the debate to the un-founded circular co-relation between subjects and objects. However, less than two centuries later, in post-modern times, another physicist named Niels Bohr would produce an essential twist to the Kantian correlational scheme where realism would become completely internalized and at the same time fully subverted. In the 20th century, Bohr would finish the job and erase the last elements of the realist {\it praxis} which Kant had still kept for his own ---essentially anti-realist--- correlational scheme, namely, the metaphysical systematization of experience through {\it categories} and {\it forms of intuition}. In tune with the positivist {\it Zeitgeist} Bohr would replace the objective-invariant account of objects of experience by the observation of single binary `clicks' in detectors and `spots' in photographic plates described in a ``commonsensical manner'' through their co-relation to macroscopic measurement apparatuses which would now even play the role ---mixing subjects with objects--- of ``material observers''. This new reference would be then supplemented by a fictional inconsistent narrative through which quantum particles would ``collapse'' in order to generate single measurement events. This procedure would be undertaken by Bohr through the addition of two {\it ad hoc} principles which helped him to support the inconsistency of the wave-particle duality (i.e., the principle of complementarity) as well as the irrepresentable bridge between the quantum microscopic realm and our macroscopic classical reality (i.e., the correspondence principle). In this way, the Danish physicist would successfully replace the guiding role that mathematical invariance and conceptual objectivity had played in modern physics by the ``common sense'' observation of classical apparatuses and measurement outcomes consequence of ---essentially irrepresentable--- quantum particles and jumps. It is this radical replacement that, as remarked by Karl Popper, would become one of the main reasons behind the triumph of instrumentalism in physics during the second half of the 20th century: 
\begin{quotation}
\noindent {\small``(a) In 1927 Niels Bohr, one of the greatest thinkers in the field of atomic physics, introduced the so-called principle of complementarity into atomic physics, which amounted to a renunciation' of the attempt to interpret atomic theory as a description of anything. Bohr pointed out that we could avoid certain contradictions (which threatened to arise between the formalism and its various interpretations) only by reminding ourselves that the formalism as such was self-consistent, and that each single case of its application (or each kind of case) remained consistent with it. {\it The contradictions only arose through the attempt to comprise within one interpretation the formalism together with more than one case, or kind of case, of its experimental application. But, as Bohr pointed out, any two of these conflicting applications were physically incapable of ever being combined in one experiment. Thus the result of every single experiment was consistent with the theory, and unambiguously laid down by it. This, he said, was all we could get.} The claim to get more, and even the hope of ever getting more, we must renounce; physics remains consistent only if we do not try to interpret, or to understand, its theories beyond (a) mastering the formalism, and (b) relating them to each of their actually realizable cases of application separately.'' \cite[pp. 100-101]{Popper63} (emphasis added)}
\end{quotation} 
In this way, physics would renounce not only its theoretical reference to {\it physis} but also to its basic {\it praxis} as guided by the attempt to produce, through mathematical invariance and conceptual objectivity, {\it moments of unity} which would allow to conceive a multiple field of experience in terms of unified theoretical (formal-conceptual) systems independently of the choice of empirical perspectives or reference frames.

\subsection{Bohr's Anti-Realism: The Non-Invariant Observability of Non-Objective `Clicks'}

Essential to the Bohrian destruction of the modern account of physics is the radical subversion and replacement of the basic concepts, principles and ideas which had, since the Ancient Greeks, determined the {\it praxis} of science for more than two millennia. Bohr, in tune with the post-modern positivist {\it Zeitgeist} would generate a completely new {\it praxis} for a new anti-realist understanding of physics where invariance, formal and conceptual consistency, objectivity as well as the systematic characterization of theoretical {\it moments of unity} would become replaced by ``common sense'' observability, contextual measurement situations, outcomes and algorithmic models supplemented by fictional narratives. We might begin our analysis by listing some of these essential steps undertaken by Bohr's effective anti-realist transmutation of physics:     
\begin{itemize}
\item {\bf The replacement of `reference frames' as referring to {\it the same} state of affairs by `bases' referring to {\it different} experimental situations or contexts.}
\item {\bf The replacement of {\it systems} and {\it states} (defined as objective-invariant {\it moments of unity}) by non-invariant and non-objective fragmentary {\it measurement outcomes}.} 
\item {\bf  The replacement of (Kantian) {\it objectivity} (understood as the condition of possibility of categorically defined moments of unity) by the {\it intersubjective communication} of binary information about `clicks' in detectors or `spots' in photographic plates.} 
\item {\bf The replacement of the {\it global} objective-invariant theoretical representation of a state of affairs by algorithmic predictive models {\it contextually} (i.e., non-invariantly) related to measurement situations and outcomes.}
\end{itemize} 

Bohr's most radical alteration of physics is to have conceived bases in the quantum formalism not as theoretical `reference frames' describing {\it the same} state of affairs form {\it different} perspectives, but as representing {\it different} experimental setups (contemporary referred to as `contexts') and consequently also {\it different} states of affairs giving rise to `singular events' that would suddenly appear in a non-causal unexplainable fashion. Indeed, Bohr's perspectivalism is not one of different viewpoints, but one of different experimental contexts co-related to non-objective (i.e., with no categorical definition of a moment unity) fragmented measurement outcomes (e.g., `clicks' in detectors and `spots' in photographic plates). As explained by John Wheeler and his student Warner Miller. 
\begin{quotation}
\noindent {\small ``What one word does most to capture the central new lesson of the quantum? `Uncertainty', so it seemed at one time; then `indeterminism'; then `complementarity'; but Bohr's final word `phenomenon' ---or, more specifically, `elementary quantum phenomenon'--- comes still closer to hitting the point.  It is the fruit of his 28 year (1927-1955) dialog with Einstein, especially as that discussion came to a head in the idealized experiment of Einstein, Podolsky and Rosen. In today's words, no elementary quantum phenomenon is a phenomenon until it is a registered (`observed' or `indelibly recorded' phenomenon), `brought to a close' by an `irreversible act of amplification'.'' \cite[p. 72]{MillerWheeler83}}
\end{quotation}
Indeed, Bohr would argue that, given a particular context, everyone witnessing the experiment would necessarily need to agree on the observed elementary quantum phenomenon. This is what, in fact, ``objectivity''  meant for him, namely, the unambiguous communication of experimental findings. What is essential to understand here is that Bohr was shifting the experimental findings from objects of experience (in the Kantian sense) to singular {\it events} which had no categorical constitution. Indeed, unlike an object, a `click' has no properties, no identity through time and its existence is strictly restricted to its appearance to a perceiving subject as a singular manifestation. Unlike objects, events have no profiles or sides and thus cannot be observed from different viewpoints or perspectives. Unlike objects, events have no identity through time and thus cannot be observed repeatedly. A `click', contrary to any theoretical {\it moment of unity}, is a perceptual singularity which turns the analysis of difference, change and motion simply impossible. Their existence, following Berkeley's famous dictum {\it esse est percipi}, can be only understood as {\it relative} to the one or many perceiving subjects ---which can be easily extended from humans to machines. In this respect, it is important to stress that such events are not ---like Bohr repeated once and again--- part of a spatiotemporal representation ---at least not like objects are in the Kantian sense. Events are not `things' which exist within space and time, but rather ``perceptual unities'' which lack completely any identity {\it through} space and time. In fact, Bohr's re-introduction of the notions of `wave' and `particle' would imply their radical re-definition ---leaving behind their metaphysical categorical definition--- in terms of `clicks'. While one single `click' would be related to the notion of `particle', many `clicks' generating an interference pattern would be linked to the notion of `wave'. This re-direction of physics to `clicks' in detectors would make possible the inconsistent redefinition by Paul Dirac of the notion of (quantum) {\it state} as related to the contextual appearance of singular observations, to quantum superpositions as linear combinations of kets and to abstract vectors with no operational content (see for a detailed discussion \cite{deRondeMassri21}). Dirac's restriction to a binary form of certainty would allow to constrain observability to single measurement outcomes erasing the reference to the intensive values which had allowed Heisenberg to develop QM in the first place. This move would be performed regardless of the non-invariance of binary values within the quantum formalism (see \cite{deRondeMassri17}). This non-invariant scheme of reasoning would become explicit in Bohr's famous reply to the EPR paper where in order to consider a subset of observables as definite valued the pre-requisite would be to specify a {\it preferred} reference frame ---i.e., a basis, or in more contemporary terms, a complete set of commuting observables. It is important to stress that such an idea of a ``preferred basis'' goes explicitly against the basic standpoint of both Greek and Modern physics. Thus, in Bohr's approach ---unlike in classical mechanics, electromagnetism and even Einstein's relativity theory--- the description of a state of affairs would become intrinsically {\it relative} to the agent's choice of a single (preferred) context (or basis).\footnote{In order to expel subjective choices it is claimed in the orthodox literature ---following Bohr--- that it is not humans but measurement  apparatuses (or nature herself) who chooses such a ``preferred basis''. This has opening the door to the famous ``preferred basis problem'' extensively discussed in the philosophical literature.} According to Bohr, changing the reference frame would mean to change the experiment and, consequently, the state of affairs itself. A basis would be then regarded, not as a formal mathematical frame to address a state form a particular abstract perspective but instead, as a system itself. This marks the abandonment of the possibility of constructing an objective conceptual scheme for QM consistent with its mathematical formalism. Objectivity in the Kantian sense, as the possibility to commonly refer to a categorically defined object of experience, would be then replaced by the mere {\it intersubjective communication} (i.e., transfer of information) of observed events between different subjects. Intersubjectivity, related now to singular measurement outcomes would not require anymore the conceptual unity presupposed within objective representations. Bohr had subverted objectivity by replacing it with intersubjectivity and simply renamed `intersubjective statements' as `objective statements'. Stressing that his account of QM was ``as objective'' as classical physics he \cite[p. 98]{D'Espagnat06} would then argue: ``The description of atomic phenomena has [...] a perfectly objective character, in the sense that no explicit reference is made to any individual observer and that therefore... no ambiguity is involved in the communication of observation.'' Bernard D'Espagnat explains this quotation in the following manner: 
\begin{quotation}
\noindent {\small ``That Bohr identified objectivity with intersubjectivity is a fact that the quotation above makes crystal clear. In view of this, one cannot fail to be surprised by the large number of his commentators, including competent ones, who merely half-agree on this, and only with ambiguous words. It seem they could not resign themselves to the ominous fact that Bohr was not a realist.'' \cite[p. 98]{D'Espagnat06}}
\end{quotation}

Applying his dualistic rhetorics, Bohr would claim that his complementarity account of physics implied a marvelous revolution that humanity was undertaking and that his interpretation of QM was as objective as classical mechanics and relativity theory. In order to make this point, Bohr confused in several occasions his perspectival interpretation of QM ---where the definite values of properties had become {\it relative} to a unique preferred reference frame (basis) or experimental content--- with Einstein's relativity theory. For example, in his Commo paper from 1929 he would write the following: 
\begin{quotation}
\noindent {\small ``While the theory of relativity reminds us of the subjective character of all physical phenomena, a character which depends essentially upon the state of motion of the observer, so does the linkage of the atomic phenomena and their observation, elucidated by the quantum theory, compel us to exercise a caution in the use of our means of expression similar to that necessity in psychological problems where we continually come upon the difficulty of demarcating the objective content.'' \cite[p. 116]{Bohr34}}
\end{quotation} 
This either deep misunderstanding or misleading explanation is also present in Bohr's famous reply to EPR: 
\begin{quotation}
\noindent {\small ``The dependence on the reference system, in relativity theory, of all readings of scales and clocks may even be compared with the essentially uncontrollable exchange of momentum or energy between the objects of measurements and all instruments defining the space-time system of reference, which in quantum theory confronts us with the situation characterized by the notion of complementarity. In fact this new feature of natural philosophy means a radical revision of our attitude as regards physical reality, which may be paralleled with the fundamental modification of all ideas regarding the absolute character of physical phenomena brought about by the general theory of relativity.'' \cite[p. 702]{Bohr35}}
\end{quotation}
Both fragments expose the deep failure of Bohr to understand  the role of  invariance in physics as one of the main pre-conditions for a (realist) ``detached subject'' representation; something which of course was contra-posed to his own (anti-realist) perspectival scheme. The analogy that Bohr attempted to make between relativity and his own complementarity account of QM is obviously wrong: relativity theory is in no way different from classical mechanics or electromagnetism when considering mathematical invariance. None of these theories reminds us of the ``subjective character of all physical phenomena'' nor ``compel us to exercise a caution in the use of our means of expression'' due to their ``difficulty of demarcating the objective content.'' On the very contrary, these theories are all objective (in the Kantian sense) and operationally invariant, allowing thus for a detached subject representation of a state of affairs. The only difference between classical mechanics and relativity is that while in the first case it is the Galilean transformations which allows us to consider all reference frames as consistently referring to {\it the same} state of affairs, in the latter case this is done through the Lorentz transformations. As Max Jammer \cite[p. 132]{Jammer74} emphasized: ``Bohr overlooked that the theory of relativity is also a theory of invariants and that, above all, its notion of `events,' such as the collision of two particles, denotes something absolute, entirely independent of the reference frame of the observer and hence logically prior to the assignment of metrical attributes.''\footnote{As Jammer \cite[p. 201]{Jammer74} continues to explain later in his book: ``[...] in Bohr's [co-]relational theory, the question `What is the position (or momentum) of a certain particle' presupposes, to be meaningful, the reference to a specified physical arrangement [...] one may formulate a theory of `perspectives', the term perspective denoting a coordinated collection of measuring instruments either in the sense of reference systems as applied in relativity or in the sense of experimental arrangements as conceived by Bohr. The important point now is to understand that although a perspective may be occupied by an observer, it also exists without such an occupancy [...] A `relativistic frame of reference' may be regarded as a geometrical or rather kinematical perspective; Bohr's `experimental arrangement' is an instrumental perspective.''} The only thing ``subjective'' here is Bohr's account of QM, in which the global consistency of the (binary) values of properties of quantum systems considered with respect to different contexts is simply abandoned.\footnote{The impossibility of a global binary valuation was later on explicitly demonstrated by the now famous Kochen-Specker theorem \cite{KS}. See also for a detailed analysis of the meaning of contextuality \cite{deRonde20, deRondeMassri17}.} 

All the operations we have mentioned above were of course perfectly aligned with Bohr's instrumentalist understanding of the mathematical formalism of QM as making reference to the instrumentalist account of measurement outcomes.
\begin{quotation}
\noindent {\small ``The entire formalism is to be considered as a tool for deriving predictions, of definite or statistical character, as regards information obtainable under experimental conditions described in classical terms and specified by means of parameters entering into the algebraic or differential equations of which the matrices or the wave-functions, respectively, are solutions. These symbols themselves, as is indicated already by the use of imaginary numbers, are not susceptible to pictorial interpretation; and even derived real functions like densities and currents are only to be regarded as expressing the probabilities for the occurrence of individual events observable under well-defined experimental conditions.'' \cite[p. 314]{Bohr48}}
\end{quotation}
Bohr stressed repeatedly that the most important lesson we should learn from QM was an epistemological one; namely, that {\it we are not only spectators, but also actors in the great drama of (quantum) existence.} 
\begin{quotation}
\noindent {\small ``Physics is to be regarded not so much as the study of something a priori given, but rather as the development of methods of ordering and surveying human experience. In this respect our task must be to account for such experience in a manner independent of individual subjective judgement and therefore objective in the sense that it can be unambiguously communicated in ordinary human language.'' \cite{Bohr60}}
\end{quotation}
The triumph of these ideas can be condensed in the orthodox widespread claim that in QM ``the properties of a system are different whether you look at them or not'' \cite{Butterfield17}. Thus, while in pre-Bohrian physics that which was considered to be {\it the same} was theoretically defined in terms of invariant-objective moments of unity, after Bohr the bearers of {\it sameness} would become the set of single non-invariant, non-objective observations of `clicks' appearing in a fragmentary, non-causal fashion within specific (preferred) experimental arrangements (or bases). As a consequence, Bohr would shift the reference of scientific research from the theoretical (formal-conceptual) invariant-objective representation of a state of affairs to the intersubjective transfer of information of non-invariant and non-objective measurement outcomes. Bohr's dogma was clear:  we are not only spectators but also actors in the great drama of quantum existence. But did all this mean that Bohr was a naive instrumentalist? A simple minded anti-realist denying the possibility to learn about a reality beyond mere observations? Well, not so fast.... Even though invariance and objectivity had been successfully replaced by Bohr's contextual intersubjective relativism ---accepted today, after Bohr, as a ``natural'' consequence of QM itself---, the Danish physicist had already re-introduced a subverted form of realism within his newly defined anti-realist program of physics...

\subsection{Bohr's Realism: Macroscopic ``Common Sense'' and the Atomist Narrative}

In order to understand the extreme effectiveness of Bohr's scheme, which continues to rule today's research in both  quantum physics and philosophy, one needs to pay special attention to the inclusion, within his own anti-realist program, of his subverted form of realism. Indeed, it is not true that Bohr can be characterized as a plain anti-realist ---as claimed by D'Espagnat and many others. Bohr would complement his anti-realist account of physics with a new characterization of realism essentially linked to fictional narratives capable, in the case of QM, of even exposing the irrepresentability of reality-in-itself. In this way, the true reference to reality would become a positive impotency. The impossibility of representation understood in positive terms would allow him not only to create an effective ``realist counterbalance'' to his instrumentalist account of measurement outcomes but also an effective trap for realists themselves who, when accepting Bohr's desiderata, would become forced to abandon their own {\it praxis} ---constrained by the formal-conceptual creation of invariant-objective systems and states. The main points of Bohr's pseudo-realism are the following:
\begin{itemize} 
\item {\bf The replacement of (metaphysical) systems of interrelated concepts by a dualistic ``realist'' reference, on the one hand, to a classical language describing our ``common sense'' macroscopic reality, and on the other, to an anti-metaphysical microscopic narrative about an essentially irrepresentable quantum reality.}
\item {\bf The introduction of {\it correspondence} acting as a principle that would secure ``bridging the gap'' between our ``commonsensical'' macroscopic intuitive account of the world and the microscopic ``irrepresentable'' quantum realm.}
\item {\bf The introduction of {\it complementarity} naturalizing the {\it inconsistent} representation of {\it the same} quantum object in terms of the dualistic self-contradicting reference to ``waves'' and ``particles''.}\footnote{Notice that, as remarked by Misak \cite[pp. 88-89]{Misak05}, Carnap's principle of tolerance played a similar role: ``The later Carnap is [...] set against the idea that we confront sentences with reality and he is led to a different brand of non-realism, with the principle of tolerance at its centre. That principle, we have seen, has it that all choices of language are to be made on `pragmatic' grounds or on grounds of utility. This views seems to entail a kind of relativism, where truth is relative to the language chosen and where there might well be two incompatible, equally good, languages.''}
\item {\bf  The  presupposition of the existence of ``quantum particles'' acting not only as the foundation of our ``commonsensical'' macroscopic (classical) reality but also as expressing the irrepresentable nature of microscopic (quantum) reality-in-itself.} 
\end{itemize}

The basic paraconsistent structure of Bohr's anti-realist realism can be found already within his famous model of the atom which marks a kernel point of inflection within the 20th century re-foundation of physics. Special attention should be given here to the notion of ``quantum particle'' playing the role of a ---supposedly--- realist presupposition. Bohr would be able to impose already in 1913 an inconsistent yet predictive algorithm ---for which he was awarded the Nobel prize--- through the brilliant introduction not only of a vague atomist narrative about tiny microscopic planetary systems but also through the addition of his famous ``quantum jumps''. Managing audience attention is the aim of all theater and the foremost requirement of all magic acts. In theatrical magic, misdirection is a form of deception in which the performer draws audience attention to one thing to distract it from another. This is the key to understand the effectiveness of Bohr's matrix. As we pointed out, in order to complete his trick Bohr did not only rely on the well known atomist images that physicists where expecting to recover, he would also tell everyone that the irrepresentable electrons orbiting an irrepresentable nucleus were capable of performing, also irrepresentable, ``quantum jumps'' which allowed them to magically disappear from their orbit and immediately reappear in another one. The story was spectacular and physicists were immediately captured. How could this happen? What were these fantastic quantum leaps? Were they actually real? How could real particles disappear and reappear at will? Where were these particles going in the meantime? The complete lack of answers did not matter. The trick had been already performed. The Danish conjurer had succeeded in drawing the focus of attention away from the critical consideration of atoms, electrons, protons and neutrons ---something that Mach had criticized just a few decades before--- to the fictional existence of ---unobserved and irrepresentable--- quantum jumps. With great confidence a young charismatic Bohr would whisper to his audience: ``It is weird because it is quantum!'' Bohr's atomist narrative, disconnected from the mathematical formalism, without a conceptual systematic characterization and no operational content, would become essential in the anti-realist discourse about microscopic reality that has pervaded up to our days. The replacement of a systematic objective-invariant theoretical representation by an algorithmic model supplemented by a fictional narrative would not only destroy the possibility of rational analysis itself, it would also reinforce the ``tolerance'' towards vagueness, inconsistency and fragmentation ---immediately redirected towards the realist praxis of creating interpretations. 

At this point it becomes essential to understand that the existence of ``quantum particles'' is the main (anti-realist) ontological presupposition present within Bohr's scheme. While in pre-Bohrian physics the {\it moment of unity} would be regarded as part of a theoretical construction explicitly defined through mathematical invariant formalisms and objective conceptual schemes capable of deriving operational testable consequences even beyond the already known experience, in post-Bohrian physics moments of unity would be considered as ``self-evident'' {\it givens} either in terms of the macroscopic observation of individual entities (e.g., tables, chairs and dogs) or in terms of measurement outcomes exposing the existence of un-observable microscopic entities (e.g., electrons, protons and neutrons). Macroscopic reality would be then justified metaphysically through the reference to an underlying microscopic reality and, vice versa, microscopic reality would be justified in observational terms as an extension of macroscopic reality. This circular reference between the microscopic and macroscopic realms would then allow to set up a self-contradictory foundation with no fundament. Depending on the needs, the debate would be then effectively redirected either to the existence (or not) of the presupposed microscopic entities as grounding ``exterior reality''  and/or ``reality-in-itself'' or to our macroscopic self-evident ``common sense'' experience of ``reality''. As a consequence, atomism would be also disconnected from its original systematic (metaphysical) architectonic and redirected to measurement outcomes. The abandonment of theoretical procedures would allow Bohr not only to claim the true existence of his own stories ---independently of theoretical  consistency and experimental support---, but ---maybe more importantly--- it would also allow him to block the possibility of any rational criticism. the basic anti-realist realist procedure developed by Bohr would impose a praxis of constant relativization of dual concepts and realms followed by the their immediate re-direction towards new dualities, relativized in order to continue a dialectic with no synthesis. The Bohrian matrix can be pictured as a highly effective M\"obius strip machine generating a fake motion through the constant creation of dualistic poles applied within a never-ending line of reasoning. Going back and forth between contradictory statements and principles, Bohr was able to create a never-ending progression of rhetorical self-justification. The Bohrian M\"obius strip of reasoning is an amazing device which forces us to remain in constant motion, always between two poles: between waves and particles, between the objective interaction of systems and subjective observations, between microscopic and macroscopic realms, between theory and measurement, between ontology and epistemology, between reality and fiction... 

\begin{center}
\includegraphics[scale=.7]{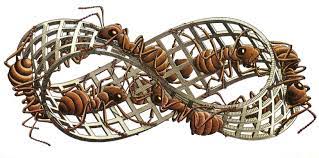} 

{\small M.C. Escher's  M\"obius strip II (1963)} 
\end{center}

Bohr's rhetorical tricks would be exposed many years later by one of his most important followers. During a meeting with Erwin Sch\"odinger which took place in Copenhagen in 1926 in order to discuss the existence of ``quantum jumps'' Bohr would apply his rhetorical powers in order to win an important battle. As Heisenberg would recall the events in his autobiography even though the many arguments that Schr\"odinger \cite[p. 73]{Heis71} had produced during the debate had allowed him to rationally conclude that ``the whole idea of quantum jumps is sheer fantasy'' the Danish illusionist, with a single move of his magic wand, would invert the burden of proof turning things completely upside-down:  
\begin{quotation}
\noindent {\small ``What you say is absolutely correct. But it does not prove that there are no quantum jumps. It only proves that we cannot imagine them, that the representational concepts with which we describe events in daily life and experiments in classical physics are inadequate when it comes to describing quantum jumps. Nor should we be surprised to find it so, seeing that the processes involved are not the objects of direct experience.'' \cite[p. 74]{Heis71}} 
\end{quotation}
Immediately after his meeting Sch\"odinger would recognize his defeat in a letter to his friend Wilhelm Wien: 
\begin{quotation}
\noindent {\small ``Bohr's [...] approach to atomic problems [...] is really remarkable. He is completely convinced that any understanding in the usual sense of the word is impossible. Therefore the conversation is almost immediately driven into philosophical questions, and soon you no longer know whether you really take the position he is attacking, or whether you really must attack the position he is defending.'' \cite[p. 228]{Moore89}} 
\end{quotation}   
Bohr's ability to deceive realists was grounded on his hollow substantialist narratives referring either to our a-systematic ``common sense'' intersubjective agreement about macroscopic reality or to an essentially irrepresentable microscopic reality composed of quantum entities that, even though would remain essentially unknowable, would anyhow play the role of grounding ontologically ``the furniture of the world''. In this complex manner realism was effectively detached from its original praxis and the general theoretical conditions which had allowed realists to guide the consistent and coherent construction of theories as formal-conceptual objective-invariant architectonics capable of accounting for a specific field of phenomena independently of subjective perspectives and reference frames. Replacing theoretical representation by inconsistent algorithms and narratives full of gaps and {\it ad hoc} rules, Bohr would establish the foundation of a new pragmatic understanding of physics which still today continues to guide contemporary research. Defying theoretical unity and consistency, inconsistent models supplemented by vague narratives would reproduce themselves fragmenting knowledge and understanding. It is in this context that atomism would end up playing a double paradoxical role, on the one hand, as grounding our ``common sense'' substantialist macroscopic understanding of the world, and on the other hand, as a fictional narrative about essentially irrepresentable microscopic entities ``approximately'' described by a set of different yet inconsistent models. The fact that physicists have become confident to accept these narratives even though they know perfectly well they are fundamentally wrong has become a trademark of contemporary physics grounded on the positivist account of physical theories in terms of purely abstract mathematical models which make contact with experience in a way that ---after more than a century--- has not been yet explained and exposes one of the deepest failures of the positivist program itself. For example, even though the reference to atoms has evaporated itself in a cloud of non-classical probabilities which even interact between each other, physicists ---and philosophers--- continue to claim that QM makes reference to a microscopic realm ---somehow implying that probabilities are ``small entities''. Exactly the same situation has taken place in the last decades with respect to the correspondence principle which during the 1980s was considered by many physicists to actually explain the quantum to classical limit. Its critical exposure during the 1990s within the young field of Philosophy of QM showing its complete failure to provide a consistent and coherent mathematical and conceptual account of this ---supposedly existent--- physical process would not stop physicists from claiming that the process did exist anyhow. In a truly Bohrian fashion physicists would simply twist the very meaning of ``solution'' and ``existence''.  Just like in Schr\"odinger's 1926 battle regarding quantum jumps, physicists would argue that none of the many philosophical exposures had proven the non-existence of the process which could be anyhow regarded to exist ``For All Practical Purposes'', in short FAPP. The creation of models which match ---a posteriori--- the collected data is then considered as a justification for an essentially irrepresentable process. the bottom line is simple: ``It works! So, shut up and calculate!''  As a direct consequence of accepting fake narratives about an irrepresentable reality, the vague reference to ``elementary particles'' has multiplied itself within contemporary physics, extending its application beyond QM to quantum field theory, string theory and ---of course--- the standard model of particle physics. As exposed in a recent article by Natalie Wolchover titled: ``What Is a Particle?'' \cite{Wolchover20}, there is no consensus within the physics community on what is the meaning or reference of such a ---supposedly--- essential concept. The meaning of the notion of `particle' fragments itself in completely different incompatible directions all of which co-exist as vague predicates of a divided, compartmented field of research. Particles are interpreted as the collapse of a quantum wave function, as the excitation of a quantum field, as the irreducible mathematical representation of a group, as vibrating strings, as the deformation of a qubit ocean, as measurement outcomes, and the list continues... In this respect, theoretical physicist at the Massachusetts Institute of Technology Xiao-Gang Wen has recognized what many students might already suspect: ``We say they are `fundamental'. But that's just a [way to say] to students, `Don't ask! I don't know the answer. It's fundamental; don't ask anymore'.'' It should not be then a surprise that John Wheeler would describe Bohr's most essential notion as ``a great smoky dragon'': 
\begin{quotation}
\noindent {\small ``The [Bohrian notion of] elementary quantum phenomenon is a great smoky dragon. The mouth of the dragon is sharp, where it bites the counter. The tail of the dragon is sharp, where the photon starts. But about what the dragon does or looks like in between we have no right to speak, either in this or in any delayed-choice experiment. We get a counter reading but we neither know nor have the right to say how it came. The elementary quantum phenomenon is the strangest thing in this strange world.'' \cite[p. 73]{MillerWheeler83}}
\end{quotation}
We believe, however, that Wheeler was essentially wrong. Bohr's anti-realist realism is not a dragon but an amphisbaena. A creature described by the Argentine Jorge Luis Borges in {\it The Book of Imaginary Beings:}  
\begin{quotation}
\noindent {\small ``Brunetto Latini's Thesaurus ---that encyclopedia which Latini, in the seventh circle of the Inferno, recommended to his former student--- is [...] straightforward: ``The Amphisbaena is a serpent with two heads, one in its meet place and the other in the creature's tail; and with both can it bite, and it runs most lightly, and its eyes gleam like live coals.'' In the seventeenth century, Sir Thomas Browne observed that there is no animal without the ``six positions of the body'' (``infra, supra, ante, retro, dextrorsum, sinistrorsum'') and he denied that the Amphisbaena could actually exist, for ``There is no inferior or former part in this animal, for the sense being placed at both extremes, does make both ends anterior.'' In Greek, ``Amphisbaena'' means ``that which goes in two directions.'' In the Antilles and in certain parts of the New World, the name is applied to a reptile commonly known as the ``double walker,'' the ``two-headed serpent,'' or the ``mother of ants.'' It is claimed that ants serve and nourish it, and also that if it is cut into two pieces, the pieces will join together again.'' \cite[p. 7]{Borges}}
\end{quotation}
\begin{center}
\includegraphics[scale=.4]{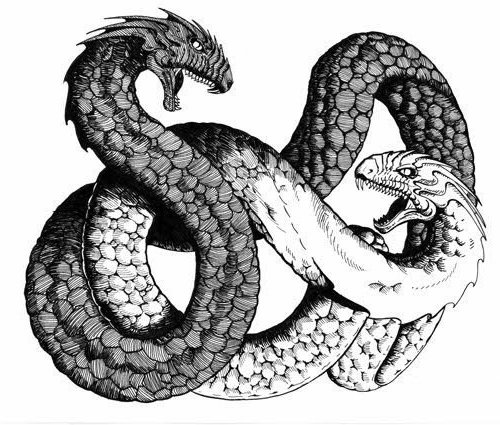}

{\small Amisphena} 
\end{center}

\section{Is a Truly Realist Account of Quantum Mechanics Even Possible?}

According to anti-realists the knowledge of subjects always begins with our perceptual ``commonsensical'' access to an ``outer-world'' composed of ``individuals'' {\it given} to us, humans, `when we open our eyes'' and are able to observe `tables', `chairs' and `dogs'. As stated in the Vienna Circle \cite{VC} manifesto: ``Everything is accessible to man; and man is the measure of all things. Here is an affinity with the Sophists, not with the Platonists; with the Epicureans, not with the Pythagoreans; with all those who stand for earthly being and the here and now.'' This empiricist standpoint is explained by Tim Maudlin in more contemporary terms: 
\begin{quotation} 
\noindent {\small ``Any empirical  science has to start from what the philosopher Wilfred Sellars called `the manifest image of the world'; that is, the world as it presents to me when I open my eyes. And of course we know that some of those appearances can be deceptive or misleading ---you know, a straw in water looks bent but it really isn't and so on ...--- but you have nowhere else to begin but with the manifest image, and then you try and produce theories that would explain it or account for it.'' \cite{Maudlin19b}} 
\end{quotation}
Given this co-relational scheme, anti-realists are forced to separate in binary terms between subjective and objective, between observable and un-observable entities, ``interior'' and ``exterior'' realities, ``concrete'' and ``abstract'', ``closed'' and ``open'', etc. This, in turn, also leads to the distinction between an intuitive ``manifest image'' of the world and an un-intuitive ``scientific image''. As remarked by Valia Allori:
\begin{quotation}
\noindent {\small ``Any fundamental physical theory is supposed to account for the world around us (the manifest image), which appears to be constituted by three-dimensional macroscopic objects with definite properties. To accomplish that, the theory will be about a given primitive ontology: entities living in three-dimensional space or in space-time. They are the fundamental building blocks of everything else, and their histories through time provide a picture of the world according to the theory (the scientific image).'' \cite[p. 60]{NeyAlbert}} 
\end{quotation}  
According to the anti-realist, there is nothing problematic in the manifest image we obtain when we observe the ``exterior reality'' that surround us as composed by ``common sense'' entities. The problem really begins when we attempt to find out the essential {\it true properties} of tables and chairs through mathematical models. Already this, marks the necessary bridge between theory and observation which the anti-realist program has completely failed to construct. But even more problematic for the anti-realist is when she attempts to model entities which are not even observable; e.g., electrons and photons. As Allori explains: ``The scientific image typically starts close to the manifest image, gradually departing from it if not successful to adequately reproduce the experimental findings. The scientific image is not necessarily close to the manifest image, because with gradual departure after gradual departure we can get pretty far away.'' Of course, the anti-realist skeptic ---who plays the role of the rational character in this drama--- is justified in doubting even about the existence of such scientific creations and might prefer to remain ``agnostic'' ---as Bas van Fraassen suggests. It is at this point that the fake realist created by anti-realists enters the scene as a fanatic preacher shouting to everyone should ---without any objective or rational justification--- they should have faith that her ``interpretation of the theory'' has finally unveiled the true ``furniture of the world''. 

Bohr anti-realist realism is of course a contemporary extension of double-thinking: a circular dialectic that generates movement without change through the generation of dualistic poles, immediately relativized and re-directed towards a new dualism... The essential modern separations between {\it res cogitans} and {\it res extensa}, the potential and the actual, epistemology and ontology, humans and nature, physics and philosophy, mater and thought... were contemporary revitalized by Bohr in terms of new dualities: microscopic and macroscopic realms, causal and non-causal interactions, rational and irrational, algorithms and outcomes, fictions and reality. According to anti-realists, science begins with the ``common sense'' access to individual entities with no explicit (metaphysical) definition whatsoever and the skeptic account of models attempting to describe them, but the impotent transcendent attempts of {\it representation} naturally fragment themselves into a multiplicity of inconsistent pictures. The ontological presupposition of individual beings becomes then a natural {\it given} impossible to describe or explicitly define. Individual entities ---observable or not--- become then ``self evident'' and ---at the same time--- un-reachable, irrepresentable. In this context, Bohr's influential principle of complementarity\footnote{The idea that \cite[p. 103]{daCostaKrause06}: ``We must, in general, be prepared to accept the fact that a complete elucidation of one and the same object may require diverse points of view which defy a unique description.''} allowed to naturalize inconsistency leading science into a dark return to pre-scientific mythical thought ---where the oppression against critical voices has become naturalized. It is by following Bohr's teachings that contemporary physicists have learned to tolerate inconsistency even within the description of one and the same entity. As Ian Hacking describes the situation: 
\begin{quotation}
\noindent {\small ``Various properties are confidently ascribed to electrons, but most of the confident properties are expressed in numerous different theories or models about which an experimenter can be rather agnostic. Even people in a team, who work on different parts of the same large experiment, may hold different mutually incompatible accounts of electrons. That is because different parts of the experiment will make different uses of electrons. Models good for calculations on one aspect of electrons will be poor for others. [...] There are lot of theories, models, approximations, pictures, formalisms, methods and so forth involving electrons, but there is no reason to suppose that the intersection of these is a theory at all.''  \cite[pp. 263-264]{Hacking83}} 
\end{quotation}
Within the scope of empirical science physics remains then impotent to make sense of reality beyond the inconsistent scrambling of ``common sense'' observations and the predictive power of models. It is at this point that the introduction of fictional narratives, interpreted by (anti-realist) scientific realists as ``mirroring'' truthfully reality-in-itself, become a fake marketing for theories which can suddenly transform themselves ---when critical questions unmask them--- ``just as a way of talking'' \cite{deRondeFM21}.  It is the combination of a skeptic (anti-realist) pragmatic scheme with a fake (realist) marketing which has become an effective way of preaching instrumentalist physics. As a result of this inconsistent scrambling, today, physicists are encouraged to perform a ``realist act'' for the vulgus in 15 minute talks where they tell to everyone amazing fictions about an unreachable world. However, once the lights are off, physicists are ready to accept according to their anti-realist skeptic rationality that all they just said did not imply any true access to reality-in-itself but rather ``just a way of talking'' that might or might not be true \cite{deRondeFM21}. 


\smallskip

Contrary to anti-realists, the fundamental cornerstone of realism is to assume the existence of {\it physis} (i.e., everything which {\it is}) as one, reality as existence in constant becoming. Everything which {\it is} must be part of reality, there is no beyond nor separation in realms, no exterior nor interior. What theories provide then is an immanent way to understand the relational being of {\it physis} through the creation of {\it moments of unity} which are able to consistently and coherently express a specific field of experience. Individuals are not {\it givens} but metaphysical creations always relative and dependent of theories. As Pauli explained: ```Understanding' probably means nothing more than having whatever ideas and concepts are needed to recognize that a great many different phenomena are part of a coherent whole.'' The notions of `wave' and `particle' are good examples of such developments in modern physics. Each of these notions creates exactly that, specific {\it moments of unity} which are capable to account for a specific field of experience. Thus, for the realist, individuals (i.e., moments of unity) are never presupposed but, on the contrary,  conceptually, mathematically and even linguistically constructioned. This was, as we have discussed above, an essential point already acknowledged by Einstein, Heisenberg and Pauli (section 1). It might be then recognized that the ontological debate which presupposes the existence of individual entities constituting the true ``furniture of the world'' should be understood as an anti-realist ``trap'' in which many contemporary realists have been caught. Since the reference to such entities is an ontological presupposition grounded on ``common sense'' and detached from theoretical representation there is no way out for the realist but to recognize her impotency to describe them in a truthfull manner.\footnote{This is the case in what has been called {\it the theory-ladenness of observation} discussed in the late 1950s and early 1960s by Norwood Russell Hanson, Thomas Kuhn, and Paul Feyerabend, there is still a pre-theoretical referential unity, an underlying $X$ which saves the co-relational scheme forged between the ``internal reality'' of subjects and the ``external reality'' of objects.} In turn, the anti-realist ontological standpoint creates the following set of problems: Are the `things' we observe truly real?  Are the macroscopic `entities' we observe grounded on un-observable microscopic `entities'? And, are the models and interpretations we provide of these entities ``useful fictions'' which help us to get along in life or ``true descriptions'' of an underlying reality-in-itself? Should we have faith in our models and interpretations or should we remain agnostic? These questions, consequence of the anti-realist standpoint and presuppositions are, for the true realist simply nonstarters. In this respect, special attention should be given to the aversion towards metaphysics ---common to both continental and analytic philosohical traditions. Bas van Fraassen \cite[p. 36]{VF02}, one of the main proponents of contemporary anti-realism, makes this point explicitly clear when he remarks: ``The story of empiricism [and anti-realism, in more general terms] is a story of recurrent rebellion against a certain systematizing and theorizing tendency in philosophy: a recurrent rebellion against the metaphysicians.'' Indeed, the ultimate triumph of anti-realism over realism during the 20th century is marked by the almost complete destruction of the general systematic metaphysical and theoretical approach which guided the realist program of science for more than two millennia. Theoretical consistency, coherency and unity have been then replaced by a modelistic approach where vagueness, inconsistency and fragmentation are tolerated as a natural consequence of the impossibility to describe a noumenic realm beyond observability. In the context of QM, where this approach was actually developed and exported to the rest of science, physicists have even learned to accept that ``nobody understands quantum mechanics'' \cite[p. 129]{Feynman67}.  

Maybe, those who feel realist inclinations might begin to wonder if there is any way out of this anti-realist maze? Can we escape the labyrinth and return to the lost realist ---Greek and Modern--- program of theoretical ---rather than ``empirical''--- science? We believe there is a very simple way out of the labyrinth given we kill the minotaur and follow the thread of Ariadna we already possess. While killing the minotaur means to destroy the pseudo-realist narratives which have replaced theoretical realism, to follow our thread implies going back to the writings and search of the last realists ---Heisenberg, Pauli, Sch\"odinger and Einstein--- leading them either to modern physics and Spinoza ---the most Greek of the modern philosophers--- or to the Ancient Greeks themselves \cite{FM20}. As realists we must commit ourselves to the realist {\it praxis} which requires us to reconsider mathematical formalisms beyond the pseudo-realist fictions that populate today's interpretational debates about the theory of quanta. In the case of QM it is only from the standpoint of a truly operational-invariant formalism already present in Heisenberg's formulation that it will be possible to derive an objective conceptual scheme which is able to account in a qualitative fashion for quantum phenomena. As we have shown in \cite{deRondeMassri19a, deRondeMassri18, deRondeMassri19b} this path is certainly open to everyone willing to re-think critically the constitution of contemporary science itself. Something that should be developed not only through the creation of formal-conceptual invariant-objective moments of unity, but also through a constant fight against vagueness, inconsistency and fragmentation. 






\end{document}